\documentclass[pra,twocolumn,showpacs,amsmath,amssymb,byrevtex]{revtex4}

\usepackage{graphicx}
\usepackage{dcolumn}
\usepackage{bm}

\begin{document}

\preprint{APS/123-QED}
\title{Non-exponential relaxation in disordered complex systems}
\author{Ekrem Ayd\i ner} \email{ekrem.aydiner@deu.edu.tr}
\affiliation{Dokuz Eyl\"{u}l University, Department of Physics,
35160 \.{I}zmir, Turkey}
\date{\today}

\begin{abstract}
We have analytically obtained the non-exponential relaxation
function for disordered complex systems applying the multi-level
jumping formalism to the fluctuation quantity which makes
diffusive motion stochastically in the disordered complex space.
It is shown that the relaxation function of disordered complex
systems decays obey to stretched exponential law.
\end{abstract}

\maketitle

Non-exponential relaxation in disordered complex systems is the
object of active research due to its implications in the
technology and in several fields of scientific knowledge. The
theoretical and experimental studies (for a recent review, see
Ref.\cite{Phllips}) show to us that the relaxation behavior in the
disordered complex systems such in glasses an supercooled liquids,
liquid crystal polymers, dielectrics, magnetic systems, amorphous
semiconductors, pinned density wave, protein dynamics, protein
folding, and population dynamics among others deviate considerably
from the exponential Debye pattern \cite{Debye}
\begin{equation} \label{1}
\Phi\left(  t\right)  =\Phi_{0}\exp\left[-t/\tau\right].
\end{equation}
and it obey to the stretched exponential relaxation function
experimentally
\begin{equation} \label{2}
\Phi\left(  t\right)  =\Phi_{0}\exp\left[-t/\tau\right]^{\alpha},
\ \ \ \ \ 0<\alpha<1
\end{equation}
often referred to as the Kohlrausch-William-Watts (KWW) function
\cite{Kohlrausch,Williams}.

The important task in the relaxation in disordered complex systems
is to extend the Debye theory of relaxation of polar molecules to
fractional dynamics, so that empirical decay functions, e.g., the
stretched exponential of KWW, may be justified. In order to
explain the origin of the non-exponential relaxation in the
disordered complex systems several models, for example,
continuous-time-random-walk models (CTRW)
\cite{Montroll,Scher,MontrollSher,WeissRubin,MetzlerKlafter,KlafterBlumen,KlafterSilbey},
random site energy models \cite{Movaghar,Bleibaum,Wichmann},
defect models
\cite{Glarum,Elliott,Klafter,Shlesinger,SherMontroll,Grassberger},
jump relaxation models \cite{Funke}, random barrier models
\cite{Kirkpatrick,Soven,Odagaki,Webman}, hierarchical models
\cite{Palmer,Kumar,Ogielski,Teitel} have been used and many
generalizations
\cite{Campbell1,Campbell2,Madrid,Vainstein,Ngai,Aydiner,Kalmykov,CoffeyKalmykov,KalmykovCrothers,CoffeyTitov}
of the Debye theory have been suggested up to now.

We remark that one of the most plausible model in among others is
definitely phase-space model in which the possible origin of
non-exponential relaxation is explained based on the idea of the
energy landscape and nontrivial energy barrier so that these
models provide a direct link between the phase-space dynamics and
slow relaxation. Indeed, several experiments and computer
simulations have been done which support the explanation of these
relaxation phenomena in the framework of energy landscape paradigm
as the result of activated diffusion through a rough energy
landscape of valleys and peaks \cite{Debenedetti,Frauenfelder}.
Therefore, it is suggested that such energetic disorder which
produces obstacles or traps which delay the motion of the particle
and introduce memory effects into the motion \cite{Shlesinger}.
Hence, in such scenario a process can be characterized by the
temporally nonlocal behavior \cite{MetzlerKlafter}.

Considering stochastic framework, in a relaxation process, the
fluctuation variable $x$ which may represent the physical quantity
such as dipole, jumps at random from one value to another with
equal probability and it takes stochastic values as $x_{1}$,
$x_{2}$, $x_{3}$,...,$x_{N}$ with time. Such a stochastic process
can be regarded as multi-level jumping process. In this study,
using multi-level jumping formalism we shall obtain relaxation
function of disordered complex system in which Brownian quantity
diffusive as stochastically. The relaxation function for a
stochastic process is given simply
\begin{equation} \label{3}
\Phi\left(  t\right)  =\Psi\left(  t=\infty\right)-\Psi\left(
t\right)
\end{equation}
where $\Psi\left( t\right)$ is the response function. The response
of the any system can be written in terms of the correlation
function as
\begin{equation} \label{4}
\Psi\left(  t\right)  =\frac{1}{k T} \left( \left\langle
x^{2}\right\rangle -\left\langle x\left( 0\right)  x\left(
t\right) \right\rangle \right)
\end{equation}
where $k$ is the Boltzmann constant, $T$ is the temperature, and
$\left\langle x^{2}\right\rangle$ is the mean-square average of
the $x$. On the other hand, $\left\langle x \left( 0\right) x
\left( t\right) \right\rangle$ is the correlation function of the
fluctuation variable $x$. The correlation function can be given in
discrete form follows
\begin{equation} \label{5}
\left\langle x\left( 0\right) x\left( t\right) \right\rangle
=\sum_{x_{0}}\sum_{x}p\left( x_{0}\right) x_{0}P\left( x_{0}\mid
x,t\right) x
\end{equation}
where the quantities $x_{0}$ and $x$ are the values of stochastic
variable  $x$ at times $0$ and $t$, respectively. Having started
at the initial state $x_{0}$ with the statistical weight $p\left(
x_{0}\right)$, $P\left( x_{0}\mid x,t\right)$ measures the
probability of propagation from $x_{0}$ to $x$ in time $t$, which
is called as conditional probability. The correlation function
measures the decay of the fluctuation variable of the physical
quantity in the system, and it is determined due to conditional
probability which depends on the nature of the stochastic system
i.e, the system is whether ordered or disordered complex.

Eqs.\,(3-5) clearly indicate that relaxation function can be
obtained from fluctuation-dissipation theory in the framework
linear-response theory if correlation function belong to the
relaxation process is known.

The conditional probability in Eq.\,(5) for disordered complex
space may be written formally \cite{HilferAnton}, but suggestive
notation, as
\begin{eqnarray} \label{6}
P\left( x_{0}\mid x,t\right) =\delta _{x0}+\frac{1}{\Gamma \left(
\alpha \right) }\int_{0}^{t}\left( t-t^{\prime }\right)
^{\alpha-1} \nonumber \\ \times \sum_{x^{\prime }}W\left(
x^{\prime }\mid x\right) P\left( x_{0}\mid x^{\prime },t^{\prime
}\right) dt^{\prime}
\end{eqnarray}
where $\alpha$ is the fractional order which plays the role of a
dynamical exponent, $\delta_{x0}$ is the initial condition at
$t=0$, and $W\left( x\mid x^{\prime}\right)$ is the jump rate for
a particle from $x$ to $x^{\prime}$. The Eq.\,(6) is known the
fractal time master equation which defines non-Markovian
stochastic process with a memory, which is a special case of the
CTRW equation \cite{HilferAnton,Hilfer}.

In the operator formalism \cite{vanKampen,Huges,Dattagupta} the
conditional probability $P\left( x_{0}\mid x,t\right)$ is simply
given with the matrix element of operator $\widehat{P}(t)$ as
\begin{equation} \label{7}
P\left( x_{0}\mid x,t\right)  \equiv\left\langle x\right\vert
\widehat {P}\left(  t\right)  \left\vert x_{0}\right\rangle.
\end{equation}
where the conditional probability satisfy
\begin{equation} \label{8}
P\left(x_{0}\mid x,0\right)  \equiv\left\langle x\right\vert
\widehat {P}\left(  0\right)  \left\vert x_{0}\right\rangle
=\delta\left( x-x_{0}\right)
\end{equation}
for $t=0$. Similarly, $\widehat{W}$ is also the jump operator
which is given by
\begin{equation} \label{9}
W\left( x^{\prime}\mid x\right)=\left\langle x\right\vert
\widehat{W}\left\vert x^{\prime}\right\rangle.
\end{equation}
Matrix representation allows to us to write the physical
quantities as an operator. Hence, the correlation function in
Eq.\,(5) can be transformed to the matrix representation
\begin{equation} \label{10}
\left\langle x \left(  0\right) x \left(  t\right) \right\rangle
=\sum_{x}p(x_{0})\left\langle x_{0}\right\vert
\widehat{X}\left\vert x_{0}\right\rangle \left\langle x\right\vert
\widehat{P}\left( t\right) \left\vert  x_{0}\right\rangle
\left\langle x\right\vert \widehat{X}\left\vert x\right\rangle
\end{equation}
where $\widehat{X}$ is the matrix representation of the
fluctuation quantity $x$, $\widehat{P}(t)$ is the matrix form of
the conditional probability, and $\left\vert ...\right\rangle$
indicates vector space of stochastic states.

The fact that $x$ takes $N$ valued stochastic variable in which
the stochastic states have been represented in $N$ dimensional
vector space. Accordingly, the fluctuation variable $x$ and
conditional probability $P\left(x_{0}\mid x,t\right)$ may be given
by the matrix representation associate a stochastic state
$\left\vert x\right\rangle$ ($x=1,2,3,...,N$ for multi-level
jumping process) with $N$ values \cite{Dattagupta}
\begin{equation} \label{11}
\left\vert 1\right\rangle =\left(
\begin{array}
[c]{c}%
1\\
0\\
0\\
.\\
.\\
.\\
0
\end{array}
\right)  ,\left\vert 2\right\rangle =\left(
\begin{array}
[c]{c}%
0\\
1\\
0\\
.\\
.\\
.\\
0
\end{array}
\right),...,\left\vert N\right\rangle =\left(
\begin{array}
[c]{c}%
0\\
0\\
0\\
.\\
.\\
.\\
1
\end{array}
\right)
\end{equation}
The stochastic states $\left\vert x\right\rangle$ (and similarly
$\left\vert x'\right\rangle$) is taken to form an orthonormal set
which provide \textit{closure} property as
\begin{equation} \label{12}
\sum_{x}\left\vert x\right\rangle \left\langle x\right\vert =1
\end{equation}
for discrete variables, and assign to it an $a priori$ occupation
probability $p(x)=\frac{1}{N}$.

Hence, the fluctuation variable $x$ can be also given by matrix
form, the eigenvalues of which correspond to stochastic variables
$x_{1}$, $x_{2}$, $x_{3}$,...,$x_{N}$, respectively. The matrix
form of the operator $\widehat{X}$ of the fluctuation variable $x$
is represented as
\begin{equation} \label{13}
\widehat{X}=\left(
\begin{array}
[c]{ccccccc}%
x_{1} & 0 & 0 &. & . & . & 0\\
0 & x_{2} & 0 &. & . & . & 0\\
0 &  0 & x_{3} &. & . & . & 0\\
. & . & . & . & . & . & .\\
. & . & . & . & . & . & .\\
. & . & . &. & . & . & \text{.}\\
0 & 0 & 0 &. & . & . & x_{N}%
\end{array}
\right)
\end{equation}

Now, after these definitions, the correlation function Eq.\,(6)
can be carried out using this formalism, however, we need the
operator form of the conditional probability in order that
calculation of it.

The time derivative of Eq.\,(5) is written as
\begin{eqnarray} \label{14}
\frac{\partial}{\partial t}P\left( x_{0}\mid x,t\right) =\frac{1}{%
\Gamma \left( \alpha \right) }\frac{\partial }{\partial
t}\int_{0}^{t}\left( t-t^{\prime }\right) ^{\alpha -1} \nonumber \\
\sum_{x^{\prime }}W\left( x^{\prime }\mid x\right) P\left(
x_{0}\mid x^{\prime },t^{\prime }\right) dt^{\prime}
\end{eqnarray}
It is seen that Eq.\,(14) contains a convolution integral with a
slowly decaying power-law Kernel $M\left(t\right)
=t^{\alpha-1}/\Gamma\left(\alpha\right)$, ensures the
non-Markovian nature of the sub-diffusion process defined by the
fractional diffusion process. This convolution integral is defined
as an operator which is known as fractional Riemann-Liouville
integro-differential operator \cite{Oldham};
\begin{equation} \label{15}
_{0}D_{t}^{1-\alpha} P\left( x_{0}\mid x,t\right)
=\frac{1}{\Gamma\left(  \alpha\right) } \frac{\partial }{\partial
t} {\displaystyle\int\limits_{0}^{t}} dt^{\prime}\frac{ P\left(
x_{0} \mid x^{\prime }, t^{\prime }\right)
}{\left(t-t^{\prime}\right)^{1-\alpha}}
\end{equation}
with $0<\alpha<1$.

Eq.\,(14) can be reduced to simply form as down
\begin{equation} \label{16}
\frac{\partial}{\partial t}P\left( x_{0}\mid x,t\right)
=_{0}D_{t}^{1-\alpha }\sum_{x^{\prime }}W\left( x^{\prime }\mid
x\right) P\left( x_{0}\mid x^{\prime },t^{\prime }\right)
\end{equation}
using Eq.\,(15). On the other hand, using operator formalism in
Eqs.\,(7) and (9), Eq.\,(16) can be rewritten completely matrix
form
\begin{equation} \label{17}
\frac{\partial} {\partial t} \left\langle x\right\vert
\widehat{P}\left(t\right)\left\vert x_{0}\right\rangle
=_{0}D_{t}^{1-\alpha } \sum_{x^{\prime}} \left\langle x\right\vert
\widehat{W}\left \vert x^{\prime}\right\rangle \left\langle
x^{\prime}\right\vert \widehat{P}\left(t\right)\left\vert x_{0}%
\right\rangle.
\end{equation}

Using closure property in Eq.\,(12),  Eq.\,(17) is reduced to
\begin{equation} \label{18}
\left\langle x\right\vert \frac{\partial \widehat{P}\left(t\right)
}{\partial t} \left\vert x_{0}\right\rangle =_{0}D_{t}^{1-\alpha }
 \left\langle x\right\vert
\widehat{W} \widehat{P}\left(  t\right)  \left\vert x_{0}%
\right\rangle.
\end{equation}
This equation can be simplify as
\begin{equation} \label{19}
\frac{\partial}{\partial t}\widehat{P}\left(t\right)=
_{0}D_{t}^{1-\alpha} \widehat{W} \widehat{P}\left(t\right).
\end{equation}
Eq.\,(19) is known fractional relaxation equation
\cite{Metzler,MetzlerKlafter}. The solution of it may be presented
in terms of Mittag-Leffler function \cite{Erdelyi}
\begin{equation} \label{20}
    \widehat{P}\left(  t\right)=E_{\alpha}\left[
    \widehat{W}t^{\alpha}\right]=\sum\limits_{j=0}^{\infty }\frac{\left( \widehat{W} t^{\alpha }\right) ^{j}}{%
\Gamma \left( 1+\alpha j\right) }.
\end{equation}
For $\alpha=1$ the Mittag-Leffler function has the standard
exponential form
\begin{equation} \label{21}
E_{\alpha =1}\left[\widehat{W}t\right] =\exp \left[
\widehat{W}t\right]
\end{equation}
whereas for $0<\alpha <1$ it interpolates the initial stretched
exponential form as
\begin{equation} \label{22}
E_{\alpha }\left[\widehat{W}t^{\alpha }\right] \backsim \exp \left[ \frac{%
\widehat{W}t^{\alpha }}{\Gamma \left( 1+\alpha \right) }\right]
\end{equation}
and, however, at the long time, the the initial stretched
exponential behavior turns over to the power-law behavior
\begin{equation} \label{23}
E_{\alpha }\left[\widehat{W}t^{\alpha }\right] \backsim
\frac{1}{\Gamma \left( 1+\alpha \right) \widehat{W}}t^{-\alpha}.
\end{equation}
In this study, we have focused the behavior of short time limit,
and we have written the conditional probability (20) for the the
simplicity as down
\begin{equation} \label{24}
  \widehat{P}\left(t\right)\equiv \exp\left[\widehat{W}
  t^{\alpha}\right]
\end{equation}
using Eq.\,(22) form. Actually, this result clearly indicates that
the character of the relaxation has stretched exponential i.e. KWW
form for the disordered complex systems. The jump matrix
$\widehat{W}$ in Eq.\,(24) may be represented in terms of
collision and unit matrix elements as
\begin{equation} \label{25}
\widehat{W}=\lambda\left( \widehat{J}-\textbf{1}\right)
\end{equation}
where $\lambda$ is the relaxation rate, $\widehat{J}$ is the
collision matrix, and $\textbf{1}$ is a unit matrix.

For the $N$-level the relaxation rate is given by
\begin{equation} \label{26}
\lambda=Nw.
\end{equation}
and the collision matrix $\widehat{J}$ is prescribed by
\begin{equation} \label{27}
\widehat{J}=\left(
\begin{array}
[c]{ccccccc}%
1/N & 1/N & . & . & . & 1/N\\
1/N & 1/N & . & . & . & 1/N\\
. & . & . & . & . & .\\
. & . & . & . & . & .\\
. & . & . & . & . & .\\
1/N & 1/N & . & . & . & 1/N%
\end{array}
\right)
\end{equation}
where $w$ is the jump rate from one value of $x$ to another.

The jump matrix $\widehat{W}$ is also given by
\begin{equation} \label{28}
\widehat{W}=\left(
\begin{array}
[c]{cccccc}%
(1-N) w & w & . & . & . & w\\
w & (1-N) w & . & . & . & w\\
. & . & . & . & . & .\\
. & . & . & . & . & .\\
. & . & . & . & . & .\\
w & w & . & . & . & (1-N) w
\end{array}\right)
\end{equation}

The usefulness of the decomposition in Eq.\,(25) is borne out by
the fact that the $\widehat{J}$ matrix has a very simple property;
it is \textit{idempotent}, i.e.,
\begin{equation} \label{29}
\widehat{J^{2}}=\widehat{J}, \ \ \widehat{J^{3}}=\widehat{J}, \ \
..., \ \ \ \widehat{J^{k}}=\widehat{J}
\end{equation}
for any integer $k>0$. This property allows us to immediately
construct the conditional probability $\widehat{P}(t)$ in matrix
form:
\begin{equation} \label{30}
\widehat{P}\left(  t\right)= \exp\left[ \lambda\left(
\widehat{J}-\textbf{1}\right)t^{\alpha} \right]
\end{equation}
Hence, using direct power series expansion and Eq.\,(29),
Eq.\,(30) can be written follow
\begin{equation} \label{31}
\widehat{P}\left(  t\right)= \exp\left(-\lambda t^{\alpha}\right)\left[\textbf{1}-\widehat{J}%
+\widehat{J}\exp\left(\lambda t^{\alpha} \right)\right].
\end{equation}

It is convenient to introduce the more general notation for basis
vectors in order to make easy of calculations. Therefore, let us
associate a stochastic state $\left\vert n\right\rangle$
($n=1,2,3,...,N$ for multi-level jumping process) instead of
$\left\vert x\right\rangle$. Thus, it is possible that Eq.\,(10)
is rewritten down form
\begin{equation} \label{32}
\left\langle x \left(  0\right) x \left(  t\right) \right\rangle
=\sum_{n,m}p_{n}\left\langle n\right\vert \widehat{X}\left\vert
n\right\rangle \left\langle m\right\vert \widehat{P}\left(
t\right) \left\vert  n\right\rangle \left\langle m\right\vert
\widehat{X}\left\vert m\right\rangle
\end{equation}
with $a priori$ occupation probability
\begin{equation} \label{33}
    p_{n}(x)=\frac{1}{N}, \ \
    \ \ n=1,2,...,N.
\end{equation}

Now, it is possible expectation values of the quantities which we
interested in;
\begin{equation} \label{34}
\left\langle n\right\vert \widehat{P}\left(  t\right)  \left\vert
m\right\rangle =\frac{1}{N}+\left( \delta_{nm}-\frac{1}{N}\right)
\exp\left( -\lambda t^{\alpha}\right)
\end{equation}
where $n,m=1,2,...,N$. The matrix of the operator $\widehat{X}$ is
\begin{equation} \label{35}
\left\langle n\right\vert \widehat{X}\left\vert m\right\rangle
=X_{n}\delta nm, \ \ \ \ \ n,m=1,2,...,N
\end{equation}
where $X_{n}$ are the allowed values of the stochastic variables.
In addition, average of the collision matrix $\widehat{J}$ is
presented as
\begin{equation} \label{36}
\left\langle n\right\vert \widehat{J}\left\vert m\right\rangle
=\frac{1}{N}, \ \ \ \ \ n,m=1,2,...,N
\end{equation}

If Eqs.\,(34) and (35) are inserted in Eq.\,(10), after a bit of
algebra, the correlation function can be written down form
\begin{equation} \label{37}
\left\langle x\left(  0\right) x\left(  t\right) \right\rangle
=\left\langle x\right\rangle ^{2}+\left( \left\langle
x^{2}\right\rangle -\left\langle x\right\rangle ^{2}\right)
\exp\left(  -\lambda t^{\alpha}\right)
\end{equation}
where the deterministic quantities are weighted averages over the
available states of the corresponding variable. The average value
of $\left\langle x\right\rangle$ and $\left\langle
x^{2}\right\rangle$ in the stationary state are given by
\begin{equation} \label{38}
\left\langle x\right\rangle =\sum_{n=1}^{N}p_{n}\left\langle
n\right\vert \widehat{X}\left\vert n\right\rangle
=\frac{1}{N}\sum_{n=1}^{N}X_{n}
\end{equation}
and
\begin{equation} \label{39}
\left\langle x^{2}\right\rangle =\sum_{n=1}^{N}p_{n}\left\langle
n\right\vert \widehat{X^{2}}\left\vert n\right\rangle
=\frac{1}{N}\sum _{n=1}^{N}X_{n}^{2}
\end{equation}
respectively.

The response function $\Psi\left( t\right)$ in Eq.\,(4) for this
process is obtained as
\begin{equation} \label{40}
\Psi \left( t\right) =\frac{1}{kT}\left\{ \left\langle
x^{2}\right\rangle -\left\langle x\right\rangle ^{2}-\left[
\left\langle x^{2}\right\rangle
-\left\langle x\right\rangle ^{2}\right] \exp \left[ -\lambda t^{\alpha }%
\right] \right\}
\end{equation}
using Eq.\,(37), on the other hand, the limit the behavior of
$\Psi\left(t\right)$ at the infinity is presented as
\begin{equation} \label{41}
\Psi \left( t=\infty \right) =\frac{1}{kT}\left( \left\langle
x^{2}\right\rangle -\left\langle x\right\rangle ^{2}\right).
\end{equation}

When Eqs.\,(40) and (41) are inserted into Eq. (3), the relaxation
function can be obtained down form
\begin{equation} \label{42}
\Phi \left( t\right) =\frac{1}{kT}\left\langle \left( \Delta
x\left( t\right) \right) ^{2}\right\rangle \exp \left[ -\lambda
t^{\alpha }\right]
\end{equation}
where $\left\langle \left( \Delta x\right) ^{2}\right\rangle $ is
the mean squared displacement which contains solely the molecular
contributions
\begin{equation} \label{43}
\left\langle \left( \Delta x\left( t\right) \right)
^{2}\right\rangle
=\left\langle x^{2}\right\rangle -\left\langle x\right\rangle ^{2}=\frac{%
2K_{\alpha }}{\Gamma \left( 1+\alpha \right) }t^{\alpha }
\end{equation}
where $K_{\alpha}$ is known the diffusion constants which is a
generalization of the Einstein-Stokes-Smoluchowski relation
\cite{MetzlerKlafter}. The $K_{\alpha}$ is defined as
\begin{equation} \label{44}
K_{\alpha}=k_{B}T/m\eta_{\alpha}
\end{equation}
where $\eta_{\alpha}$ is the friction coefficient which is a
measure for interaction of the particle with its environment, and
$m$ denotes mass of the particle
\cite{MetzlerKlafter,MetzlerNonnenmacher,MetzlerBarkai,Barkai}.

Finally the relaxation function in Eq.\,(42) can be simplified as
\begin{equation} \label{45}
\Phi \left( t\right) =\frac{1}{kT}\frac{2K_{\alpha }}{\Gamma
\left( 1+\alpha \right) }t^{\alpha }\exp \left[ -\lambda t^{\alpha
}\right].
\end{equation}
The prefactor in Eq.\,(45) can be taken $\Phi_{0}$ which is a
constant in Eq.\,(2). As a result, the relaxation function (45) is
consistent with Eq.\,(2), which clearly indicates that fluctuation
quantity in a disordered complex system decays with time as
stretched exponential i.e, KWW form.

In this study, applying the multi-level jumping formalism to the
fluctuation quantity which make diffusive motion stochastically in
the disordered complex system, we have analytically obtained the
relaxation function KWW form in terms of correlation function in
absence of the external field. It is concluded that multi-level
jumping process formalism is quite powerful technique for the
modelling of the Brownian motion to obtain the relaxation function
of disordered complex systems.


\end{document}